# TRACEABILITY AS AN INTEGRAL PART OF SUPPLY CHAIN LOGISTICS MANAGEMENT: AN ANALYTICAL REVIEW

*Dharmendra K. Mishra[1], Sebastien Henry[2], Aicha Sekhari[1], Yacine Ouzrout[1]*
[1] University Lyon 2 Lumiere, DISP Laboratory, France
[2] University Claude Bernard Lyon 1, DISP Laboratory, France

**ABSTRACT**

**Purpose**: Supply chain has become very complex today. There are multiple stakeholders at various points. All these stakeholders need to collaborate with each other in multiple directions for its effective and efficient management. The manufacturers need proper information and data about the product location, its processing history, raw materials, etc at each point so as to control the production process. Companies need to develop global and effective strategies to sustain in the competitive market. Logistics helps companies to implement the effectiveness across the business supply chain from source to destination to achieve their strategic goals. Traceability has become one of the integrated parts of the supply chain logistics management that track and trace the product's information in upstream and downstream at each point. All the stakeholders in the supply chain have different objectives to implement the traceability solution that depends on their role (e.g. manufacturers, distributers or wholesalers have their own role). The information generated and needed from all these actors are also different and are to be shared among all to run the supply chain smoothly. But the stakeholders don't want to share all the information with other stake holders which is a major challenge to be addressed in current traceability solutions. The purpose of this research is to conduct thorough study of the literatures available on the traceability solutions, finding the gaps and recommending our views to enhance the collaborations among the stakeholders in every point of the business supply chain. The study will be helpful for further researchers to propose a traceability meta-model to address the gaps.

**Design/methodology/approach**: The relevant literatures with keywords supply chain management, traceability, logistics management are searched from the trusted database. The scope and the objectives of the research is set based on which the papers are filtered as per the titles and abstract. Proper analyses of the screened papers are done and the recommendations are given.

**Findings**: Traceability solution is popular among the industries since a long for their supply chain management. After proper analysis of the available literatures about the traceability solution the industries use, this research will identify the gaps based on which we give the proper recommendations and perspectives of our work.

## Introduction

Success of company depends on its ability to compete in global market. The competitive markets require that business supply chains are highly agile, effective and efficient (Helo p. Et al., 2014). To achieve the above goal, company need to determine the status (Identity, location, processing history, physical status etc) of the product in supply chain. Every stakeholder either the manufacturing team, distributer, supplier, retailer or the users in the business supply chain must know the products information throughout its life cycle. Traceability is a system companies use to know the status of the product in supply chain. The global traceability system is among them which manufacturers use in order to control their production processes. According to Webster's Dictionary, "Traceability is the ability to follow or study out in detail, or step by step, the history of a certain activity or a process". Thus product's traceability can be defined as the ability to follow the product's data and information from its conception to end in the supply chain. Traceability helps the industries in optimizing supply chain, knowing market status, improving product's quality etc. It also makes consumers safer by providing detailed information about where an item comes from, what its components and origin are and about its processing history (Regattieri A et al., 2007). The EU legislation to keep traceability system makes food industries mandatory to implement the traceability system. But to keep the traceability system just to satisfy the legal requirements will be a financial burden for the companies. It is necessary to make aware the companies that traceability is something more that helps them to sustain in the market. It helps not only to control the production process but also to make the market strategy, set dynamic pricing and quality control. There must be a proper balance between order processing, inventory management, manufacturing, warehousing and distributing to sustain in the competitive environment. Logistics management helps the companies to meet these requirements for which every actor must know the product's data and information at every point in supply chain. The supply chain is complex today and thus there should be an efficient system to integrate the whole production process that trace and track the product's and process information in the business supply chain. And thus traceability is an integral part of the supply chain logistics. In this paper we analyse the literatures related with logistics and traceability and give our recommendation based on the research gap that will be helpful for further researchers to develop and implement such system.

**Traceability**

The supply chain has become more complex as the products move up and down more frequently from/to various points. There are many actors involved in the product supply chain and hold lots of information and data. The industries need to maintain product's quality to protect their brand as well as to sustain in the competitive market. A sudden recalling of product can save people's life when a product which are directly related with their health, are found contaminated or with low quality standard. In this case, the manufacturers must identify the origin of the problem and location of the product. These are some issues which demand the visibility of the products in whole supply chain and traceability is a system which supports to address all these issues.Traceability is the ability to trace the history application or location of an entity by means of recorded identification throughout the supply chain (ISO 8402:1995).
Traceability is often discussed as two linked process:
1. **Track:** Ability to physically locate articles or items inside a facility - to a specific location or to identify articles or items used to fulfill an outbound sales order *(e.g., where it is and where it went).*It is the process of finding the product in downstream in the supply chain.

2. **Trace:** Ability to search historical records identifying manufacturing processes and the source of ingredients or components, etc. *(e.g., how it was processed and what was done).* It is the process of finding the products data in upstream in the supply chain.

Traceability is also discussed in terms of internal and external traceability. In the internal traceability, the traceability partner receives one or more traceable items that are processed internally to create another one or more traceable item(s) (GS1). The internal process may include processing raw materials, packaging, storing in the warehouse, or destroying. External traceability is the term used to describe the flow of the product in supply chain. Here, a traceability partner receives a traceable item and handed over to one or more traceability partner(s) (GS1). That is, external traceability tracks the location of the product in the supply chain during its physical flow. Traceability is also defined by considering three parameter space(s), time (t) and volume (v) (Samarasinghe R., 2009). By space it is meant that the traceability system finds the location of the products and process. Similarly by the term t means when the information is traced and finally v reflects the volume of the information traced.

**Benefits of traceability**

To have the product's information from its conception to end is the key to success of any company and traceability system is a best tool to access and disseminate the information. It helps every actors in the supply chain to add value on the production and distribution system from planning to disposal of the product and thus to achieve innovation in product designing process. Some of the key benefits of the traceability system are described below.

1. Transparency: The actors in the supply chain need to know the product's know how status to plan and act. Users of the product must have to know all the details of the product like ingredients of the product, processing history, date of manufacture and country of origin etc. The product manufacturing process from conception to end should be transparent so that the actors access these information whenever needed. Traceability helps to maintain the transparency throughout the supply chain.

2. Quality Control: Customer satisfaction is key for the company to sustain in the market. The customer satisfies when they have confidence in using the product and can get every information about the product when they need. Company need to adopt zero tolerance in the quality control mechanism during the product manufacturing process. Whenever a quality related issue is brought into the notice of any actors in any point of supply chain, a proper and immediate action need to be taken to improve the design and manufacturing process so that proper remedy action can be taken in the next lot or batch of the production. For this, the actors have to trace back in the supply chain and have to find at which point and what errors in the production has been occurred. A traceability system with trace back capability helps the actors to manage the quality control process.

3. Reduce time to market: Traceability system tracks all the related information on every point in the supply chain creating a bond between every departments of a company from order processing through inventory management, processing, packaging, warehousing and despatching. This information helps the actors to act on time so that all the ordered products are manufactured and sent to market on time. This helps in overall cost reduction of the production process and thus increase the profit of the company.

4. Fighting Product Counterfeiting and protecting brand: A company gains brand through years of experience of designing innovative product, maintaining the quality and thus by winning the customer satisfaction. This brand will be collapsed in a second when some fraud products with same brand having low quality appear into the market. Traceability helps to track the original product in the supply chain and thus helps the actors in fighting product counterfeiting.

5. Improvement in SCM: Traceability system helps to improve the efficiency of supply chain management process by reducing the cost mainly logistics, providing all the information from product conception to retail in the market (Bosana T. and Gebresenbet G., 2013). This improvement increases the collaboration among the supply chain actors and thus develops their economic and technical competence.

6. Increase intactness with consumers: As discussed before customer's satisfaction is key to business success and the satisfaction is achieved when we reach to the customers. More the product information we provide to customers more will they close to the product and manufacturers. With a proper traceability system, consumers can access the product related information anytime they need and thus help actors to intact with them.

7. Increase competence: To know customer's buying behaviour for a company is necessary to manufacture the product in needed time and launch accordingly in the market. For a retailer the traceability database allows him to know which product is sold in which quantity and in which season. Similarly he also knows about the type and brand of the product sold and thus he can accordingly decide what volume of specific product he should order in which time so as to meet the customer's requirements. This helps him to increase his confidence to compete with the competitors. It also may be the source of competitive advantage of supply chain partners because the traceability system helps to increase efficiency of SCM by solving the product's safety problems, enabling the company to understand its logistic system and enabling them to produce the quality products on time (Bosana T. and Gebresenbet G., 2013).

The above are some of the benefits company can take by implementing the traceability system. In the next section we present the analysis of related works.

## Related Works

A company witness a huge financial and brand loss if it does not able to deliver the product on right time in the market. Proper production control process not only helps to manage the order but also helps the company to innovate the product designing process and logistics traceability is best tool to achieve this. Automatic traceability system has great impact on the supply chain from reducing the profit loss to the quality control (Sahin E., Dallery Y. Et al. 2002). In their work, the authors opine that traceability system helps the actors to analyse the customer behaviour and gives the proper knowledge of out of stock situation (Sahin E., Dallery Y. Et al. 2002). In their work X. Wang and D. Li opines that to have the traceability system to just meet the legal requirement will add extra cost burden to companies (Wang X. and Li D. 2006). And thus they discus to integrate the traceability system with supply chain process. The traceable data can be used to innovate the business process to achieve and implement better production control and inventory management. The authors also opine that the benefits of the traceability system include production efficiency through overall quality improvement, promotion management and dynamic pricing and improvement of supply chain logistics and distribution process (Wang X. and Li D. 2006). In another work, a RFID based traceability system is implemented to manage the logistics process of aftermarket automotive parts distribution of a Tibetan company (Liu C.S., Trappey C.V. et al 2008). In this work the author compared the as-is model with the to be model of the logistics process from make order through make invoice, pick and sort goods etc to transportation and proved that the proposed traceability model reduces the time spent on these process by 60%. J. Zhang, P. Fenget et al identified the requirements of traceability system and proposed EPC/RFID based traceability process and architecture model (Whang J., Feng P. Et al. 2008). The supply chain is complex today and as such every actors must act effectively to make the supply chain efficient. R. Samarasinghe et al in their work identified that due to the complexity of supply chain, there should be an efficient system to integrate the whole production process and to trace all the product and process information to maintain the quality and thus they opined that traceability is an integrated part of supply chain management (Samarasinghe R., Nishantha G.G.D et al 2009). In this work, authors implemented a central traceability system where an interface is made to connect the stakeholders enabling the total traceability that is vertical and horizontal traceability. In phase one, a solution was proposed for the SME's where the paper records are scanned and put in the local data base which is then connected with database of the main company for the purpose further tracing the production process (Samarasinghe R., Nishantha G.G.D et al 2009). A traceability system based on EPC global network was developed to collect RFID data and RFID events to trace and record the logistics information (Minbo L and Chen C. 2009). Many works suggest that RFID is important and efficient tools to develop the traceability system. An RFID system consists of an antenna and a transceiver, which read the radio frequency and transfer the information to a processing device, and a transponder, or tag, which is an integrated circuit containing the RF circuitry and information to be transmitted. The required product's information are stored in the RFID tag which are traceable by the RFID reader at any point in the supply chain within 10 meters range. Unlike barcodes, there does not need to be direct line of sight between the tag and reader. An integrated logistics based on RFID and barcode for the dairy products was proposed by Y. Zhang et al. Here the information from raw material gathering, processing, packing, storing and distributing is traced and stored in the database for further access by the stakeholders in supply chain from product manufacturer to consumers (Zhang Y. and Wang L 2009). People are more concerned to know the quality information specially for the food product before using it. So a proper system must be implemented that will provide all the information to the users on the requested time. Bosona T et al in their work opines that traceability is an integral part of logistics management and propose a new definition of food traceability stating that it is the part of logistics management that capture, store and distribute the adequate information (Bosona T., Gebresenbet G. et al 2013). In this work the author analyzed the driving force, benefits and barriers of implementing food traceability system and suggested that chain traceability is required which is the part of integrated logistics system. They also suggested that further researches have to be done on the sector of technological aspects, link between traceability system and production system, standardization of information exchange, awareness creation and the efficiency of traceability system (Bosona T., Gebresenbet G. et al 2013). The general impression is that traceability is highly beneficial for the optimization of production process and it will be more beneficial if this system is integrated with logistics management due to the complex nature of the supply chain.

**Analysis and Practical Issues**

The literatures presented in the above section mainly focus on developing the traceability system is beneficial to the company and suggest developing such system based on RFID. There are very few works that have discussed about the complexity issue of supply chain. No work addressed to manage such issue.

The supply chain is very complex today. There are multiple layers of actors both horizontally and vertically. A manufacturer is producing more than one product and at the same time it has more than one supplier for the same products or the parts of the product. Fig 1 below shows the multiple layers of the supply chain.

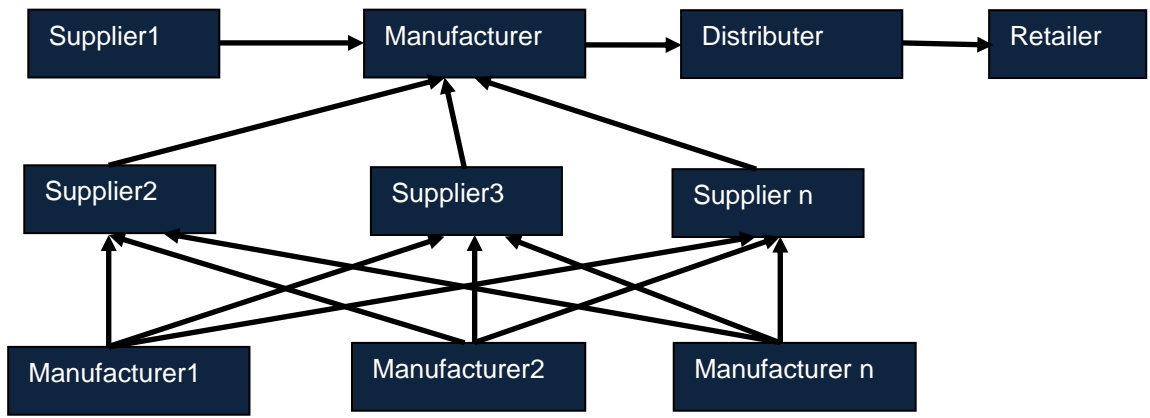

Fig. 1: Complex Supply chain

As shown in the figure, there is more than one manufacturer giving the parts or the product to the same supplier. Likewise there is more than one supplier supplying the same products or its parts to one manufacturer. This complexity could go further deeper in two directions. From the traceability point of view, these actors have two types of data viz. private and public. Public data is shared among all the actors but they don't want to share the private data globally. Fig 2 below shows the interrelation between the data of the multiple trading partners in 2 levels. To make supply chain effective, efficient and agile, the data and information of the product at every point of the supply chain must be traceable. For example, if a product is found faulty in any point of supply chain from initiation to the use phase, it must be needed to identify the causes of the problem for which actors must supply all the related information either it is private or public. In the global supply chain, where trading partners are dispersed across the globe may not know each other and in this case it does not give some information for example processing history of the product to other partner. But this information must be needed to identify the cause of the problem. This is a major challenge in current supply chain management process. In order to make the supply chain agile the actors must act in collaboration but there are two reasons which make actors not to act in collaboration. First is , there is a level of mistrust between the actors because every actors don't know every other in the global supply chain and the second is all the actors have their own traceability system and thus there will be data alignment problem. There is a problem of the information extraction between the actors. A common understanding is needed among the actors to use same standard for their traceability system but it is really difficult to make it possible. Similarly, there exists some SMEs in the supply chain which don't have enough budget to implement the IT system. They are still using the paper based traceability information and in this case again there will be the problem of information extraction.

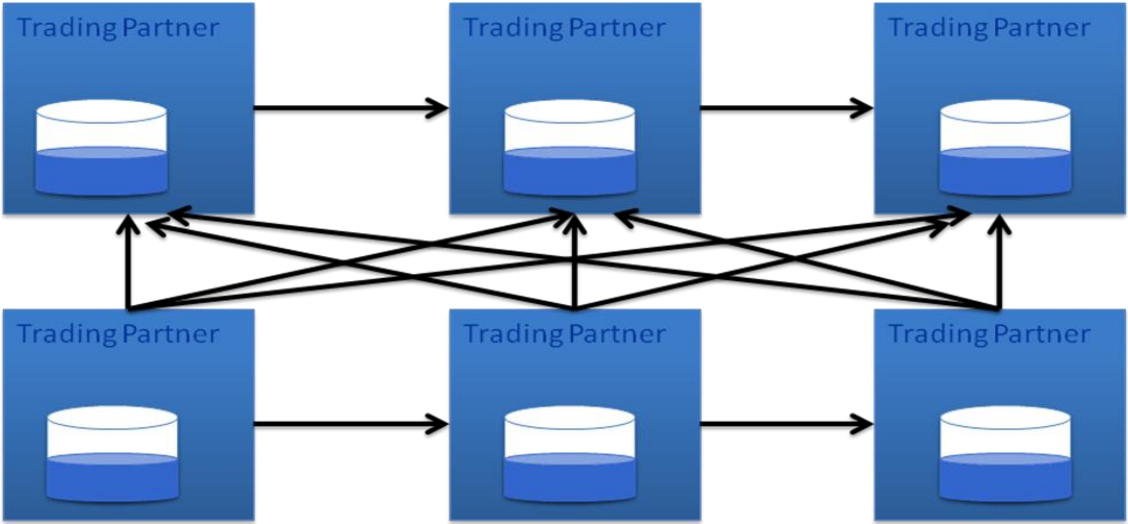

Fig 2: Information exchange model

Based on the above discussion we find three issues which are needed to be solved.

1. How to make collaborations among multi-stakeholders in supply chain?
2. How to align the Information systems among the stakeholders?
3. How to effectively store the data to maintain privacy and security (Location Transparency)?

A proper model must have to be designed to address the above stated issues. Fig 3 shows an integrated traceability model.

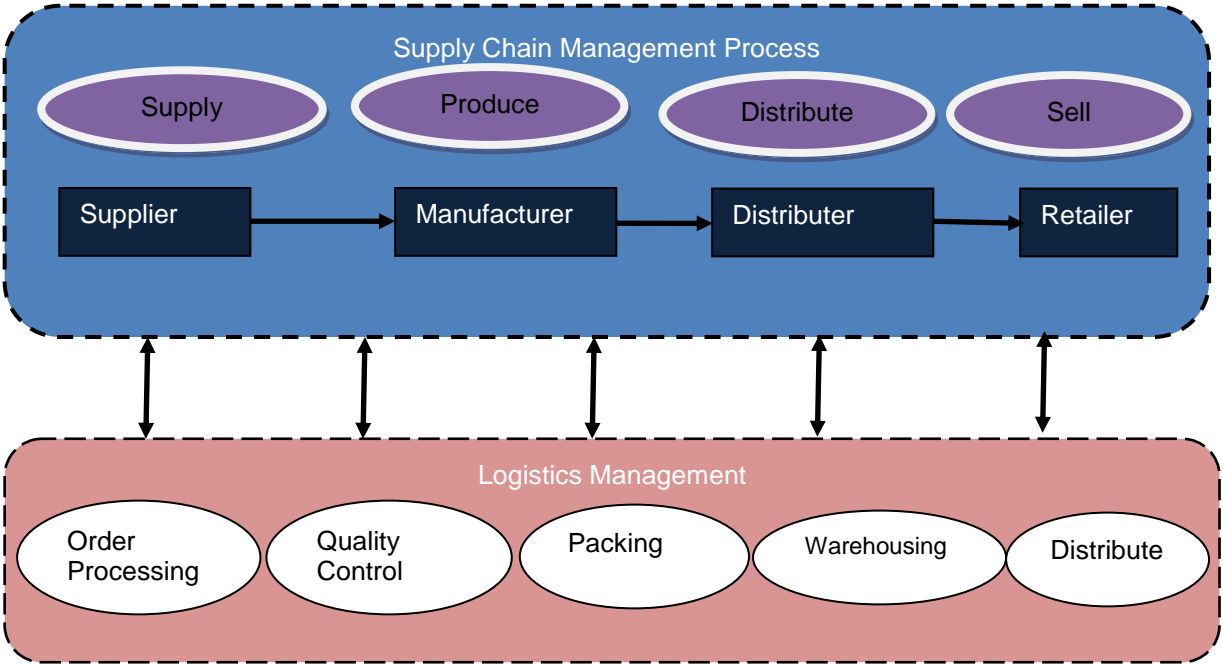

Fig. An integrated traceability Model

An integrated traceability model with logistics management have to incorporated because company's success depends not only on just to have the products data but also it is needed to use those data in the logistics management which help them to manage all the process from order management through processing, quality control, packaging, warehousing, to distributing. This is a key for the company to sustain in the market and accordingly succeed.

**Conclusion**

Growing competitive markets demand the companies to have their supply chain management process agile, effective and efficient. To achieve this companies need the product's data and information throughout the supply chain. An automatic traceability system allows them to track the location of product in downstream and trace the processing history and other treatment of the product in upstream in supply chain. But the supply chain is complex today and all the actors are dispersed geographically around the globe. They don't know and trust each other and thus don't want to share some crucial information globally. This is a major challenge of implementing an efficient traceability solution. In this research we analyse the existing literatures about traceability system in the supply chain management process. Identify the major issues to be addressed to make the solution effective and finally we propose an integrated model with logistics management process to manage the overall production control mechanism. This work is useful for the further researchers to develop some model to address the issues discussed.